\documentclass[
reprint,
floatfix,
aps,
prb,
amsmath,
twocolumn,
superscriptaddress,
amssymb,
tightenlines,
]{revtex4-1}

\pdfoutput=1
\usepackage{amsfonts,amssymb}
\usepackage[subnum]{cases}
\usepackage{mathrsfs}
\usepackage{amsmath}
\usepackage[none]{hyphenat}
\usepackage{hyperref} 
\usepackage{bm} 
\usepackage{natbib}
\usepackage{soul} 
\usepackage{color}
\usepackage{longtable}
\usepackage{ textcomp }
\usepackage{verbatim}
\usepackage{fancyhdr} 
\usepackage{lastpage} 
\usepackage{extramarks} 
\usepackage{graphicx} 
\usepackage{lipsum} 
\usepackage{ amssymb }
\usepackage{slashed}
\usepackage{tikz}
\usepackage{comment}
\usepackage[caption=false,listofformat=parens, subrefformat=parens]{subfig}
\usepackage{ marvosym }
\usepackage{array}

\usepackage{amsthm} 

\let\baraccent=\= 
\renewcommand{\=}[1]{\stackrel{#1}{=}} 

\theoremstyle{definition}

\theoremstyle{remark}

\newcolumntype{C}[1]{>{\centering\let\newline\\\arraybackslash\hspace{0pt}}m{#1}}

\begin{document}
\title{Giant violation of Wiedemann-Franz law in doping layers of modern AlGaAs heterostructures}

\author{M. Sammon} 
\email[Corresponding author: ]{sammo017@umn.edu} 
\affiliation{School of Physics and Astronomy, University of Minnesota, Minneapolis, MN 55455, USA}
\author{Mitali Banerjee}
\affiliation{Department of Physics, Columbia University, New York, NY, USA}
\author{B. I. Shklovskii} 
\affiliation{School of Physics and Astronomy, University of Minnesota, Minneapolis, MN 55455, USA}

\received{\today}

\begin{abstract}
We analyze the data of the recent paper Nature \textbf{559}, 205 (2018) and show that it contains an observation of thermal and electric conductivities of the doping layers in GaAs/AlGaAs heterostructures which violates the  Wiedemann-Franz law. Namely, the measured thermal conductivity of the doping layers is similar to that of a metal while the electrical conductivity is exponentially small. We conjecture that these results are related to the exciton contribution to thermal conductivity calculated in several recent theoretical works for metallic samples.
\end{abstract}

\maketitle

\maketitle

In the recent paper\cite{Moty(2018)} the quantization of the thermal conductance in a two-dimensional electron gas (2DEG) in units of $\kappa_0 T$ ($\kappa_0\equiv\pi^2 k_B^{2}/3h$, where $k_B$ is the Boltzmann constant and $h$ is the Planck constant) was comprehensively studied in the Quantum Hall regime at temperatures $T\sim20$ mK. The 2DEG resided in a GaAs quantum well and was provided by two Si doping layers separated from the 2DEG by AlGaAs spacers of equal thickness. Although the authors were focused on transport properties of the 2DEG, in this note we concentrate on the results obtained for transport properties of the doping layers. Samples with two different designs of doping layers were used. Both designs exhibited quantization of the thermal conductance within the 2DEG, with very different thermal conductance of the doping layers. In the structure with an old-fashioned design of the doping layers where Si donors are located in AlGaAs, the authors observed no parallel thermal conductance from the doping layer. On the other hand, in the largest mobility modern samples with short period superlattice (SPSL) doping layers, they observed substantial additional (parallel) thermal conductance $\sim 3.3\kappa_0 T$ from the doping layer. This is a large, ``metallic", thermal conductance. If the Wiedemann-Franz (WF) law were valid this would correspond to an electrical conductivity $\sigma = 3.3 e^2/h$ of the two doping layers.

\begin{figure}
	\includegraphics[width=\linewidth]{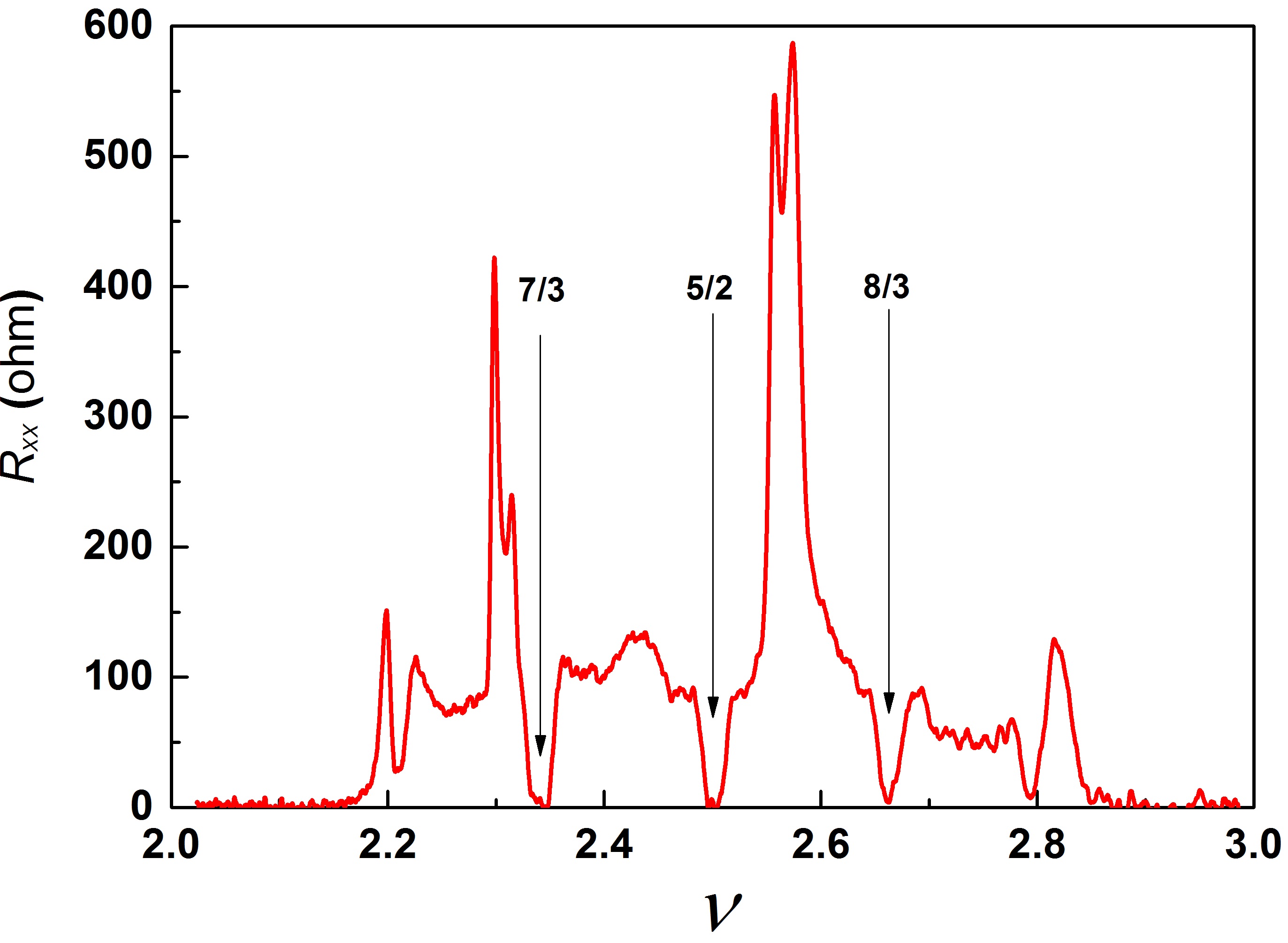}
	\caption{Longitudinal resistance $R_{xx}$ vs. filling factor $\nu$ for the SPSL sample of Ref.\,[\onlinecite{Moty(2018)}]. This trace was taken at $T=$ 10 mK. At $\nu=2$, we see that $R_{xx}< 3$ $\Omega$.  }\label{fig:Rxx}
\end{figure}

Now we try to estimate the parallel electrical conductivity $\sigma$ of the two SPSL layers, using the Hall bar $R_{xx}$ data taken at 10 mK shown in Fig.\,\ref{fig:Rxx}, borrowed from the supplementary material of Ref.\,[\onlinecite{Moty(2018)}]. 
We concentrate on the $R_{xx}$ minimum at $\nu=2$, but similar results can be obtained for the minimum at $\nu=3$. Because the doping layers and the 2DEG share all contacts, their conductivity tensors are added when considering the measured conductivity of the combined system of the 2DEG and doping layers. This results in 
\begin{equation}\label{eq:rho}
\rho_{xx} \simeq \sigma/\sigma_{xy}^{2},
\end{equation}
for the longitudinal resistivity $\rho_{xx}$ of the combined system.
Here $\sigma_{xy}$ is the Hall conductivity of the 2DEG, and we have neglected the $\sigma_{xx}$  of the 2DEG, which should be very small for such a temperature and the corresponding magnetic field $B\simeq 6$ T at $\nu=2$ . The length of the Hall bar in the $x$-direction is twice larger than the width in the $y$-direction, so that  $\rho_{xx} = R_{xx}/2$. We see from Fig.\,\ref{fig:Rxx} that $R_{xx} < 3$ $\Omega$, thus we take $\rho_{xx} < 1.5$ $\Omega$ as a good estimate. Then Eq.\,(\ref{eq:rho}) leads to the inequality for the conductivity of the doping layers  $\sigma < 2\times 10^{-4} e^2/h$, which should be compared with $\sigma = 3.3 e^2/h$ obtained above by assuming the validity of the WF law. Thus, the SPSL doping layers have a thermal conductivity that is at least four orders of magnitude larger than what the WF law predicts.

 SPSL doping layers are designed in a way that donors keep a fraction of their electrons (see Refs.\,[\onlinecite{SammonZudov}] and [\onlinecite{SammonTianran}] and references therein). These excess electrons are situated in narrow AlAs wells which surround a narrow GaAs layer doped by Si donors in the middle. Each excess electron is localized in the vicinity of a donor in a so called compact dipole. In view of the large effective mass of AlAs electrons are localized, although the concentration of donors is so large that the localization length can be larger than the average distance between donors in the GaAs layer. Most likely the electric conductivity of such a layer obeys Efros-Shklovskii (ES) variable range hopping law\cite{ES_1975,footnote1} and should be very small at $T=20$ mK.\cite{SammonZudov,SammonTianran} At the same time, electrons choose their host donors to minimize their energy and screen the random potential of the charged donors. This is why SPSL structures have very high mobility. This should be contrasted with the direct doping design, where the Si donors form the so called DX centers\cite{Mooney1990_DX} which tend to freeze electrons in the doping layers at relatively high temperatures. As we argue below, the large localization length of the SPSL doping layer is a possible explanation of the observed metallic thermal conductivity.

Such anomalously fast energy transport was conjectured when it became clear that in the quantum Hall effect the prefactor $\sigma_0$ of the ES hopping law $\sigma_{xx}=\sigma_0\exp[-(T_{ES}/T)^{1/2}]$ observed in the minima of $\rho_{xx}$ is universally of the order of $e^2/h$.\cite{Polyakov1993}  It was conjectured that the ES law hopping there is electron-electron interaction assisted. Indeed, the standard phonon assisted hopping would lead to a square of the small electron-phonon interaction constant in $\sigma_0$ which is apparently absent.

 It was later conjectured\cite{Berkovits1999} that neutral delocalized excitations providing energy for electron hops are electron-hole pairs (excitons). It was emphasized that in spite of the Coulomb gap in the one particle density of states, the electron-hole density of states is approximately constant near zero energy,\cite{ES_1975} which should help exciton delocalization.

The theory of the additional contribution of soft exciton modes to the thermal conductivity of disordered metallic 2DEGs which do not obey the WF law was developed in Refs.\,[\onlinecite{Savona2004,Smith2005,Catelani2005,Catelani2007,SchwietePRB,SchwieteJETP}]. The authors showed that at the lowest temperatures the thermal conductivity of these soft exciton modes is of the order of  $\kappa_0 T$. We conjecture that this thermal conductivity can be used even when the electron states are localized, but the localization length is larger than the average distance between electrons and the conductivity is near $e^2/h$. In this case the standard electron contribution to the thermal conductivity which obeys the WF law becomes exponentially small, and therefore the contribution of soft modes dominates. This allows us to explain the experimental results\cite{Moty(2018)} discussed above.

Here we talked only about electrons whose localization length is larger than the average distance between electrons. In the ``lightly doped" disordered electron system, where the localization length is much smaller than the average distance between electrons\cite{Gutman2016} thermal conductivity of 2DEG is predicted to be proportional to $T^{3}$. Apparently SPSL layers of Ref.\,[\onlinecite{Moty(2018)}] are more heavily doped and have a thermal conductivity of the order of $\kappa_0 T$.

$\phantom{}$
\vspace*{2ex} \par \noindent

	We are grateful	to M.\,Heiblum, V.\,Umansky, I.\,V.\,Gornyi, A.\,Kamenev, and M.\,A.\,Zudov for helpful discussions. M.\,Sammon was supported primarily by the NSF through the University of Minnesota MRSEC under Award No. DMR-1420013

\end{document}